\documentclass[aps,prl,amsmath, amssymb, english, ' twocolumn,reprint,floatfix, superscriptaddress]{revtex4-2}

\usepackage{multirow}
\usepackage{float}
\usepackage[normalem]{ulem}
\usepackage{textgreek}
\usepackage{amssymb}

\usepackage{bbm}
\usepackage{comment}
\usepackage{soul}

\usepackage{tikz}
\usepackage{microtype}
\usepackage{tkz-euclide}
\usepackage{tkz-graph}
\usepackage{epstopdf}
\usepackage{epsfig} 
\usepackage[applemac]{inputenx} 
\usepackage{amsmath}

\usepackage[unicode]{hyperref}
\usepackage{orcidlink}
\hypersetup{
 colorlinks,
 citecolor=blue,
 filecolor=blue,
 linkcolor=blue,
 urlcolor=blue
}

\newcommand{\ket}[1]{\left|#1\right>}
\newcommand{\bra}[1]{\left<#1\right|}

\graphicspath{{./Images/}}

\date{}
\begin{document}

\title{Optically Tunable Spin Transport in Bilayer Altermagnetic Mott Insulators}
\author{Niklas Sicheler \orcidlink{0009-0001-7398-5547}}
\affiliation{Institut f\"ur Theoretische Physik und Astrophysik and W\"urzburg-Dresden Cluster of Excellence ct.qmat, Universit\"at W\"urzburg, 97074 W\"urzburg, Germany}
\author{Roberto Raimondi \orcidlink{https://orcid.org/0000-0001-5174-6759}}
\affiliation{Dipartimento di Matematica e Fisica, Universit\'a Roma Tre, Via della Vasca Navale 84, 00146 Roma, Italy}
\author{Giorgio Sangiovanni \orcidlink{https://orcid.org/0000-0003-2218-2901}}
\affiliation{Institut f\"ur Theoretische Physik und Astrophysik and W\"urzburg-Dresden Cluster of Excellence ct.qmat, Universit\"at W\"urzburg, 97074 W\"urzburg, Germany}
\author{Lorenzo Del Re \orcidlink{0000-0002-8765-2866}} 
\email{lorenzo.re@uni-wuerzburg.de}
\affiliation{Institut f\"ur Theoretische Physik und Astrophysik and W\"urzburg-Dresden Cluster of Excellence ct.qmat, Universit\"at W\"urzburg, 97074 W\"urzburg, Germany}
\date{\today} 


\begin{abstract}
Altermagnets are a novel class of materials that combine antiferromagnetic spin ordering with non-relativistic spin splitting (NRSS) in their band structure, making them promising candidates for spintronics applications without requiring strong spin-orbit coupling. In this work, we investigate a two-dimensional bilayer Mott insulator that exhibits altermagnetic order. The interplay between spin and layer degrees of freedom gives rise to a complex symmetry-breaking pattern involving both magnetic and interlayer-coherent components. A key control parameter in the system is the layer polarization, which can be tuned via an external gate voltage. We show that applying an in-plane electric field with opposite signs in the two layers induces a polarization current that drives a spin current in each layer. While the polarization current is isotropic, the resulting spin current exhibits strong anisotropy and can be reversed by adjusting the photon energy. These findings suggest new avenues for manipulating spin transport in altermagnetic systems via electric and optical means.
\end{abstract}
\maketitle

\noindent
{\bfseries \emph{Introduction} --} Altermagnetism has recently emerged as a new paradigm in condensed matter physics \cite{kunes2019,Mazin2021,Smejkal2022_a,Smejkal2022_b,McClarty2024,Jian2024}, where a many-body system simultaneously displays antiferromagnetic properties--determined by its spin arrangement in real space--and ferromagnetic-like features, characterized by non-relativistic spin splitting (NRSS) in its band structure. The NRSS arises because different sublattices are not related by a simple translation combined with a spin-flip operation; instead, additional symmetry operations, such as rotations or mirror reflections, are involved.  The absence of net magnetization combined with spin-split bands makes altermagnets good candidates for antiferromagnetic spintronics applications \cite{Baltz2018} without requiring materials with strong spin-orbit coupling. In insulating altermagnetic states, the spectrum of magnetic excitations (magnons) is expected to split into chiral-up and chiral-down branches \cite{Smejkal2024,PhysRevLett.133.156702}, a property that could be exploited to generate magnon currents via the spin-Seebeck and Nernst effects \cite{Cheng2016,Zyuzin2016,Mook2019,naka2019} in response to a temperature gradient \cite{Cui2023}.
Lately, a great interest in two-dimensional altermagnetism \cite{PhysRevLett.133.086503,PhysRevLett.132.236701,khan2025altermagnetism,Parshukov2025,Wang2025,daghofer2025altermagneticpolarons,lanzini2025,rao2025van} has emerged, with a growing focus on bi-layer systems \cite{Zeng2024,Pan2024,Hodt2024,Qi2024,Zeng2025,parshukov2025exotic}, that can be engineered for technological and sensing purposes.

In this Letter, we study a model Hamiltonian describing two-dimensional bilayer Mott insulators that host altermagnetic order. Layer degrees of freedom open new avenues for spontaneous symmetry breaking, giving rise to ordered phases involving not only spin observables but also interlayer coherence. As a result, the emerging order is not purely magnetic, akin to phenomena observed in multi-orbital systems~\cite{ikeda2012,chubukov2016,Hayami_2016,mazzone2022}, and features interlayer excitonic condensation. A key observable in our system is the layer polarization \( P_z \), which can be controlled in bilayer platforms via an external gate voltage. We demonstrate that an in-plane electric field with opposite signs in the two layers induces a polarization current, which in turn generates a spin current in each layer. While the dipole response remains isotropic, the spin conductivity is highly anisotropic. Crucially, this spin response to a polarization (counterflow) current vanishes in a collinear antiferromagnetic excitonic insulator, but is symmetry-allowed and finite in the altermagnetic counterpart. Remarkably, the sign of the layer-resolved spin current can be reversed by tuning the photon energy.
\begin{figure}
    \centering
    \includegraphics[width=1.\columnwidth]{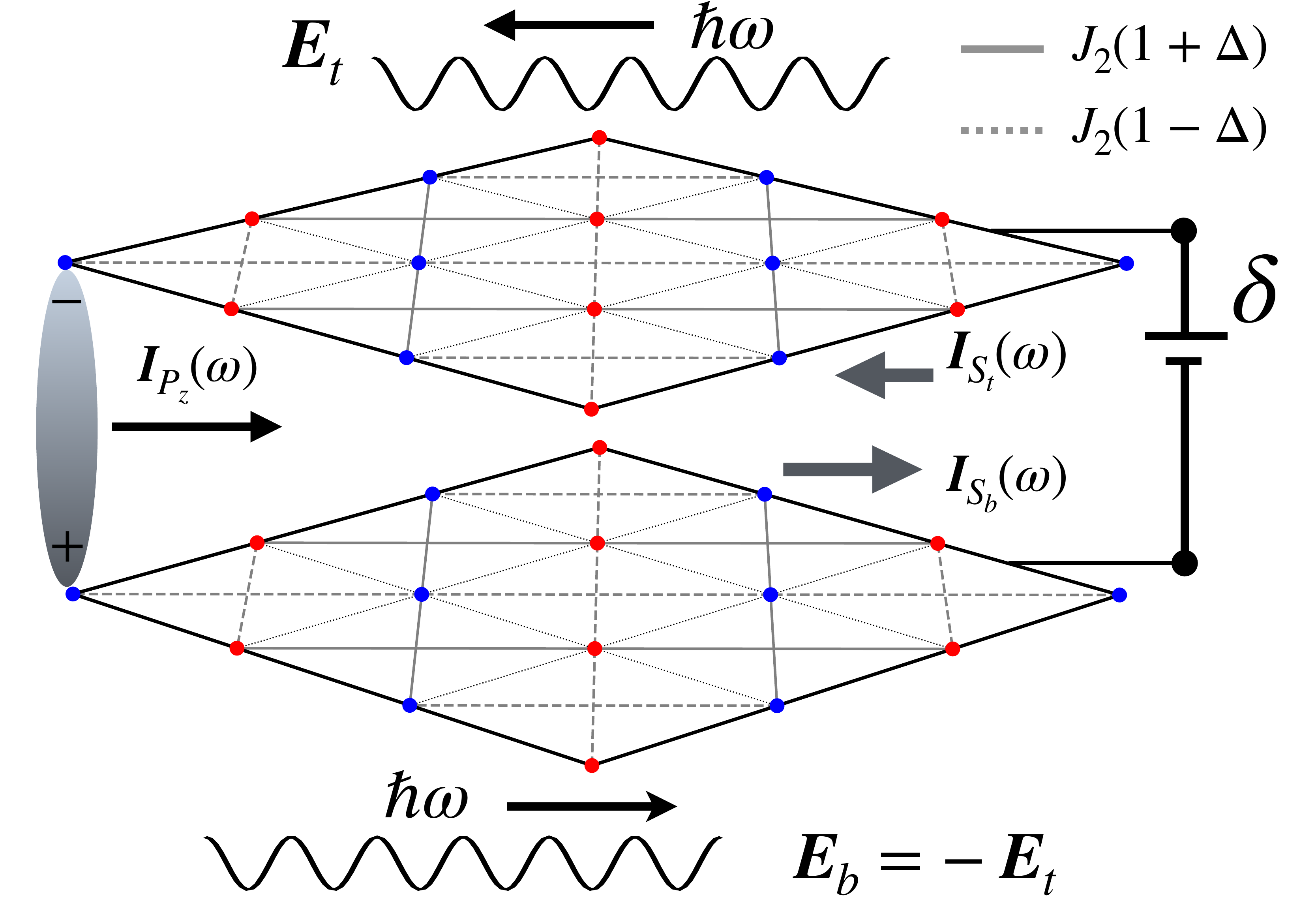}
    \caption{\textbf{Sketch of the bilayer system.} Layer polarization $P_z = n_{t} -n_{b}$, where $t$ ($b$) refers to the top (bottom) layer, is controlled by a gate voltage $\delta$. Blue (red) sites indicate the two spin sublattices.
   The second-neighbor exchange interaction is also shown; 
    it explicitly breaks translational invariance when $\Delta \neq 0$, leading to altermagnetic order. When two counter-propagating electric fields are applied to the layers in opposite directions, they couple to the dipole charges (i.e., the difference in occupation between the two layers). This, in turn, couples to the spin degrees of freedom and generates a spin current $I_S$ when $\Delta \neq 0$. The sign and intensity of $I_S$ depend on the driving frequency $\omega$, which can reverse the relative direction of the spin current between the two layers.
 }
    \label{fig:sketch}
\end{figure}
\\ \par 
\noindent
{\bfseries \emph{The model} --} We consider the square-lattice Hubbard model with four local degrees of freedom per site, labeled as $\alpha = {\ket{1},\ket{2},\ket{3},\ket{4}}$. For specificity, these states are associated with spin ($\uparrow,\downarrow$) and layer (top, bottom) indices: $\ket{1} = \ket{\uparrow \rm{t}}$, $\ket{2} = \ket{\downarrow \rm{t}}$, $\ket{3} = \ket{\uparrow \rm{b}}$, and $\ket{4} = \ket{\downarrow \rm{b}}$. In the strong coupling regime and at integer fillings $n = 1$ or $n = 3$, the low-energy physics of the Hubbard model is effectively captured by a Kugel-Khomskii-type spin model \cite{Kugel_1982}, described by

\begin{align}\label{eq:Heis} 
    \hat{H}& = {J_1}\sum_{i}\sum_{\tau \,\in\,  \text{n.n.}}\hat{\mathcal{P}}_{i,i+\tau} + J_2\sum_i\sum_{\rho = \pm d_1} {[1+(-1)^{i} \Delta]}\hat{\mathcal{P}}_{i,i+\rho} \nonumber \\
    & + J_2\sum_{i}\sum_{\rho = \pm d_2} {[1 - (-1)^i\Delta]}\hat{\mathcal{P}}_{i,i+\rho} + {\delta}\sum_i \hat{P}^z_i,
\end{align}
where $\hat{\mathcal{P}}_{ij}= \sum_{\alpha\beta}S^\alpha_\beta(i)S^\beta_\alpha(j)$, with $S^\alpha_\beta = \ket{\alpha}\bra{\beta}$ denoting the SU(4) spin operators. The exchange couplings $J_{ij}$ include both nearest-neighbor (NN)  and next-nearest-neighbor (NNN) contributions with amplitudes $J_1$ and $J_2$, respectively.   Furthermore, we consider an extra field $\Delta$ adding up to the usual NNN coupling which distinguishes between  processes along the two diagonals $d_1 = (1,1)$ and $d_2 = (1,-1)$ in an alternate fashion \cite{Das2024,delre2024dirac,liu2024} doubling the unit cell of the original lattice [see Figure \ref{fig:sketch}].
The second term in Eq.(\ref{eq:Heis}) introduces an energy imbalance between the layers through the polarization operator $\hat{P}^z = \ket{1}\bra{1} + \ket{2}\bra{2} - \ket{3}\bra{3} -\ket{4}\bra{4}$, favoring the bottom layer for positive $\delta$. Throughout this work, we assume $\delta > 0$ without loss of generality. The Hamiltonian remains structurally unchanged under a relabeling of the states $\ket{1},\ldots,\ket{4}$, indicating that it can equivalently describe systems with orbital imbalance or a Zeeman field. In the SU(4)-symmetric case $\delta = 0$, all four local states are equally populated. As $\delta$ increases, the system gradually reduces to an effective SU(2) model in the large-$\delta$ limit, where one layer becomes entirely unoccupied and the remaining layer is half-filled.

To characterize various order parameters and observables, we define the operators
$\hat{O}^{ab} = \sum_{\alpha\beta} (\tau_a \otimes \sigma_b)_{\alpha\beta} S^{\alpha}_{\beta}$,
constructed from the generators of SU(4). Here, \( a, b = 0,1,2,3 \), with \( \sigma_0 \) and \( \tau_0 \) denoting the \( 2\times2 \) identity matrix, and \( \sigma_{1,2,3} \), \( \tau_{1,2,3} \) the Pauli matrices acting on the spin and layer (or pseudo-spin) degrees of freedom, respectively.
In this representation, spin operators associated with the top and bottom layers can be expressed as $\hat{S}^{k}_{t(b)} = \frac{1}{2}[\hat{O}^{0,k} \pm \hat{O}^{3,k}]$ for $k = 1,2,3$. The inter-layer coherent processes underlying excitonic ordering are captured by the eight operators $\hat{O}^{a,b}$ with $a = 1,2$ and $b = 0,1,2,3$.
\\ 

\noindent
{\bfseries \emph{The Ground State} --}
\begin{figure}[htbp]
    \centering
   \includegraphics[width=0.49\columnwidth]{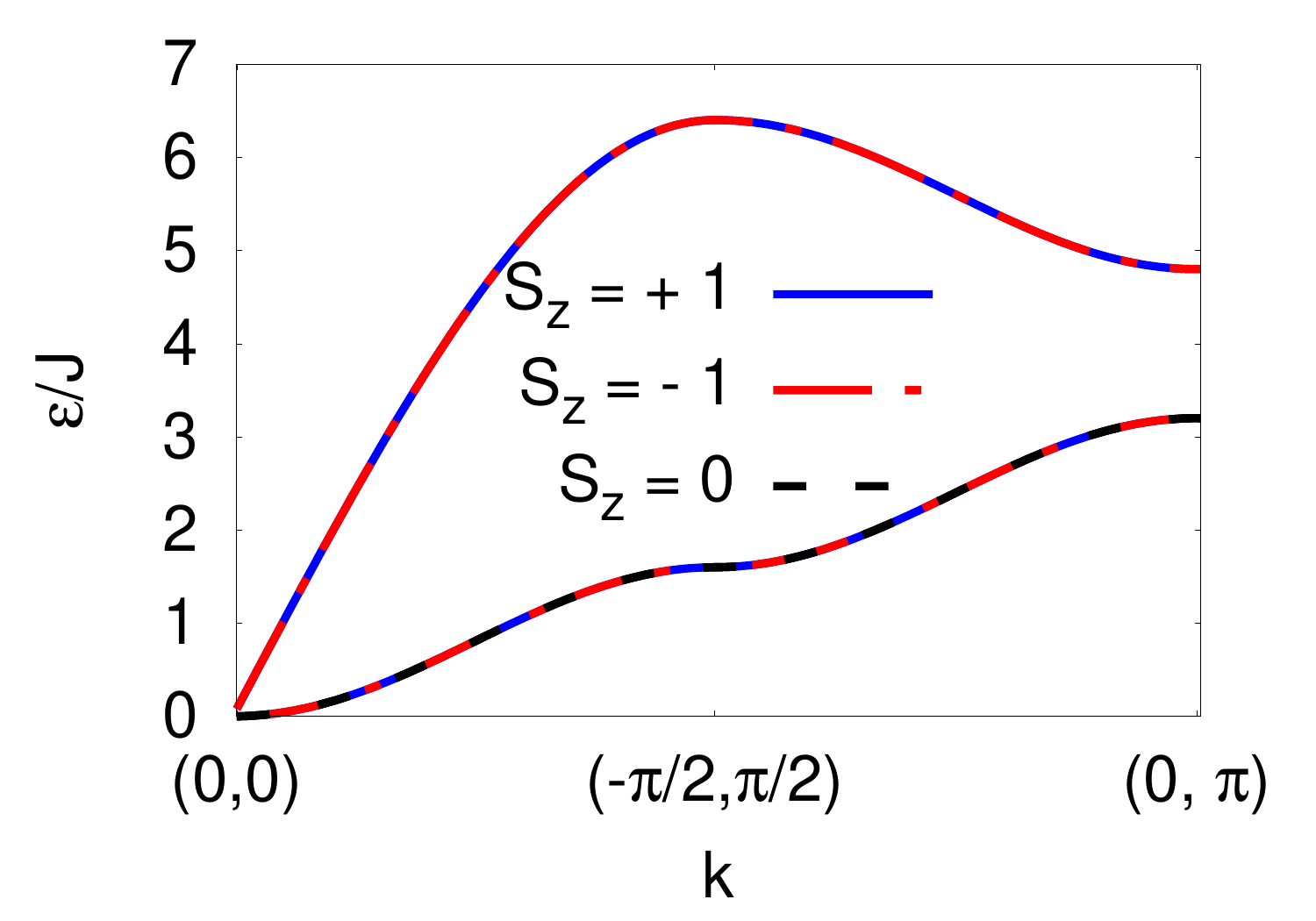}
   \includegraphics[width=0.49\columnwidth]{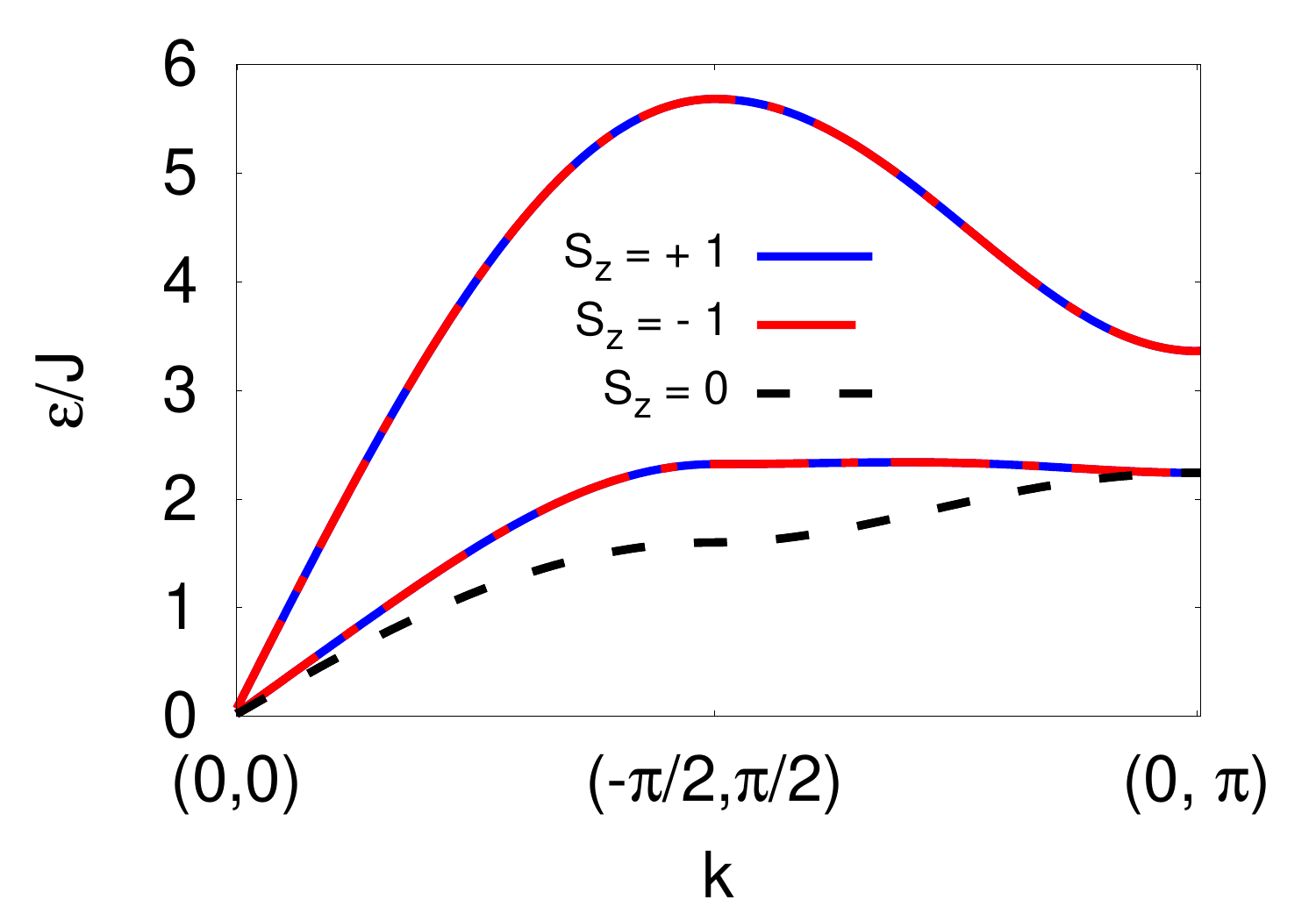}
   \caption{
Spectrum of excitations above the antiferromagnetic  excitonic ordered state for polarization values. In this case we set $\Delta = 0$, so the system does not display altermagnetism.
(Left panel) $P_z = 1$: antiferromagnetic phase;
(Right  panel) $P_z = 0.7$: antiferromagnetic + excitonic phase;
}
    \label{fig:flavour_waves_AF}
\end{figure}
We employ mean-field theory to compute the state with the minimum energy of the system.  
To derive a mean-field solution for arbitrary $\delta$ within these limits, we use a product-state Ansatz $\ket{\Psi} = \prod_i \ket{\psi_i}_i$ and minimize the classical energy $E_{\text{cl}} = \bra{\Psi} H \ket{\Psi}$ \cite{Joshi1999}. Specifically, we choose an ansatz with in-plane spin order and a homogeneous layer polarization $P_z$, given by
$\ket{\psi_i} =  \sqrt{\frac{{1+P_z}}{2}}\left(\frac{1 + e^{i Q_s^t\cdot R_i}}{2}\ket{1} + \frac{1 - e^{i Q_s^t\cdot R_i}}{2}\ket{2}\right) 
    +e^{i Q_p\cdot R_i}\sqrt{\frac{{1-P_z}}{2}}
    \left(\frac{1 + e^{i Q_s^b\cdot R_i}}{2}\ket{3} + \frac{1 - e^{i Q_s^b\cdot R_i}}{2}\ket{4}\right)$
where $Q_p$, $Q_s^b$, and $Q_s^t$ represent the wave-vectors corresponding to the relative modulations of the pseudo-spin (layer), bottom spin, and top spin, respectively \cite{PhysRevResearch.6.023082}.
At \( J_2 = 0 \), we observe an extensive ground-state degeneracy for all values of \( P_z \) except in the fully polarized case.
However, a small but finite \( J_2 \) lifts this degeneracy when \( P_z \) is nonzero. In this regime, for \( J_2/J_1 < 1/2 \), the classical energy is minimized by \( Q_s^t = Q_s^b = (\pi, \pi) \), indicating a collinear antiferromagnetic ordering in both layers, and by \( Q_p = (\pi, 0) \) or \( (0, \pi) \), signaling a canted stripe ordering of the layer pseudo-spin.
We refer to this state as the N\'eel excitonic state.

At $J_2/J_1 = 1/2$, we observe a one-dimensional degeneracy similar to the one of the SU(2)-limit, where $Q_s^t=(\pi,s-\pi),Q_s^b=(\pi,\pi-s),Q_p=(\pi,s)$ with $0<s<\pi$.
For $J_2/J_1 > 0.5$, the optimal spin wave-vectors are given by $Q^t_{s} = Q^b_{s} = (\pi,0)$ and $ Q_p= (s,\pi)$.

Quite remarkably, for all values of \( J_2/J_1 \) and \( P_z \), the classical energy is independent of the altermagnetic field \( \Delta \), which cancels out once the energy is averaged using our factorized ansatz.

In the next sections we will focus on the N\'eel excitonic phase that we expect to be stable for low enough values of second-neighbor exchange that we fix to $J_2=0.2\,J_1$ and for high enough values of the layer polarization. 

\begin{figure*}
    \centering
    \includegraphics[width=.75\columnwidth]{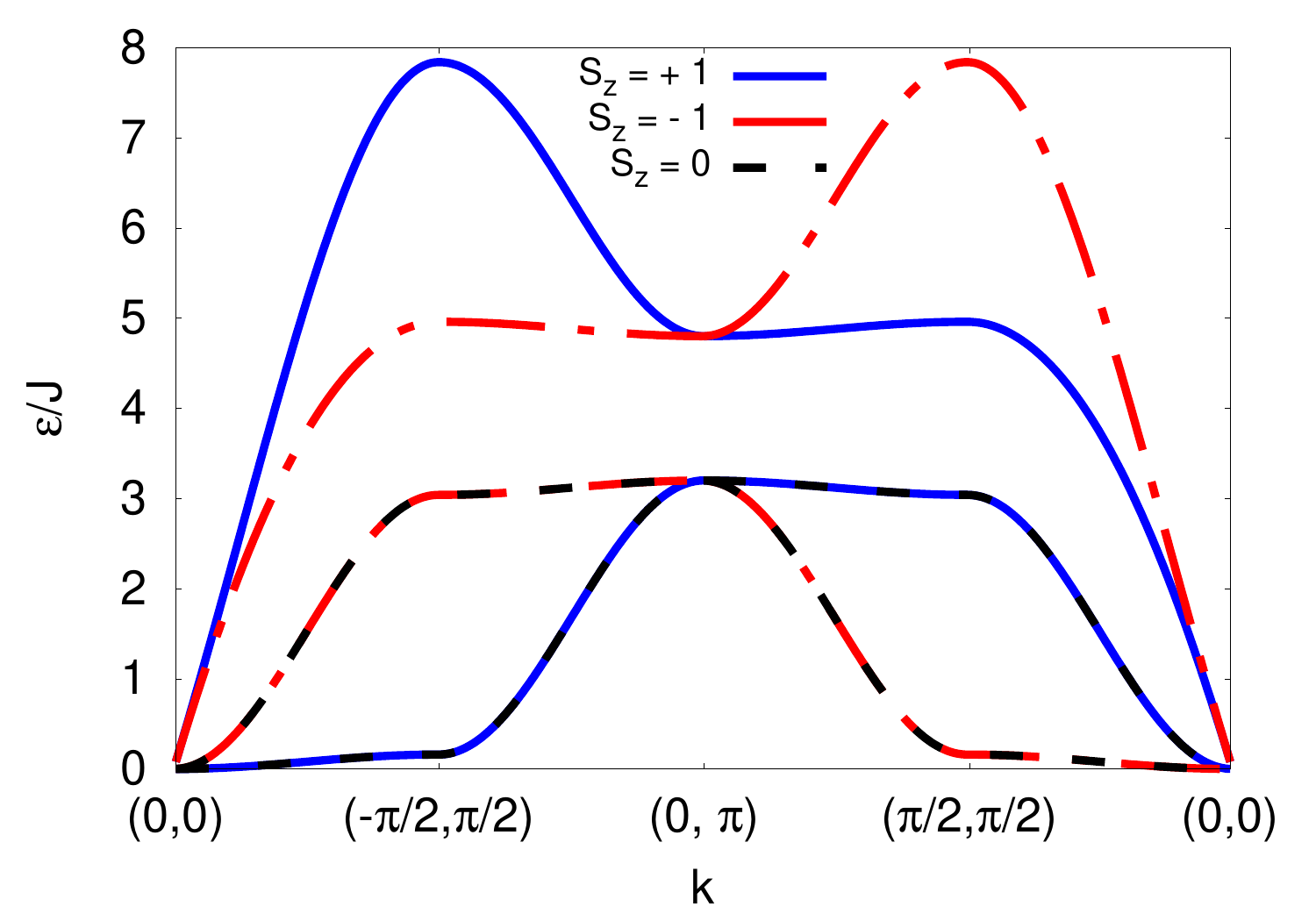}(a)
    \includegraphics[width=0.75\columnwidth]{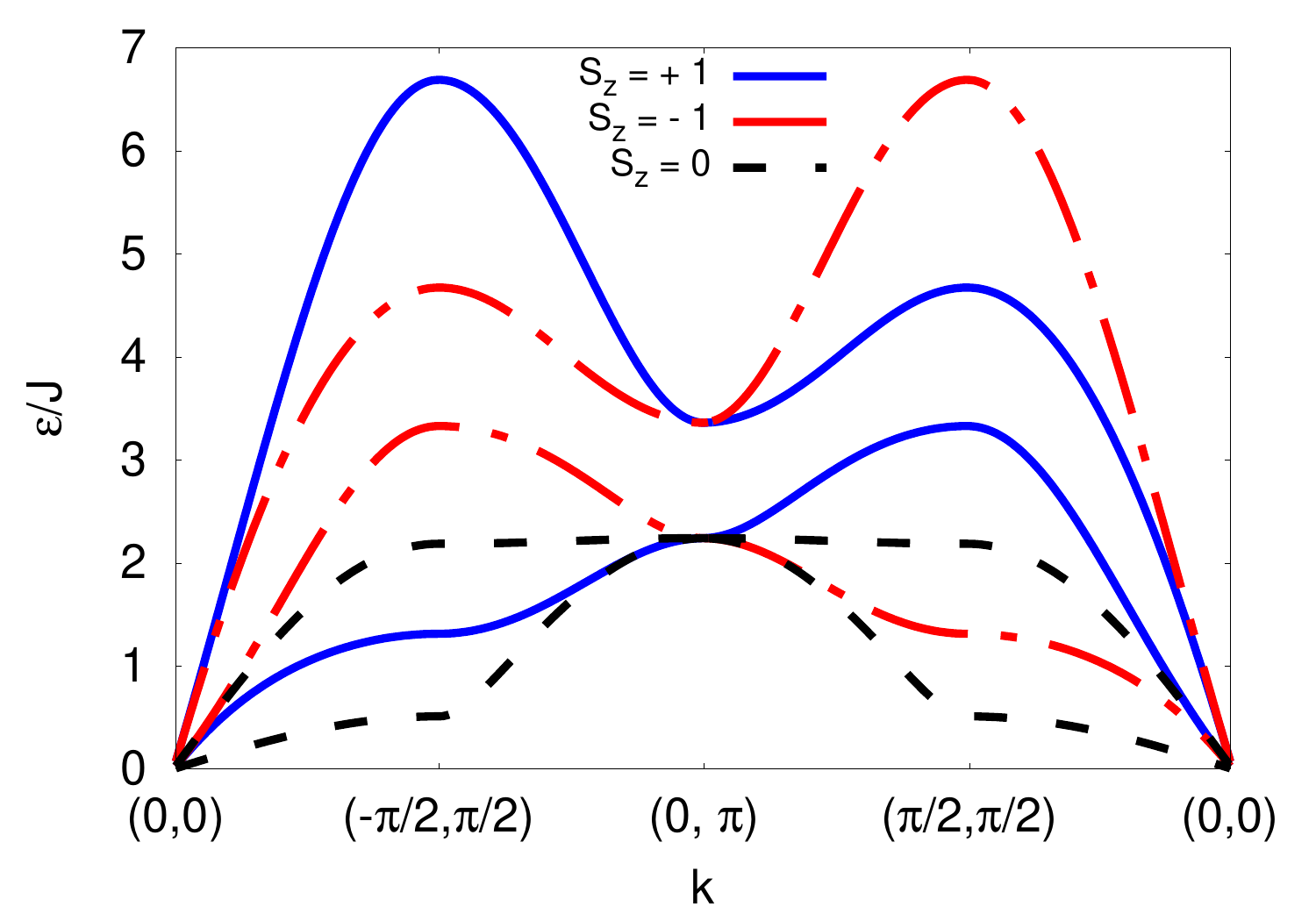}(b)
   \caption{
Spectrum of excitations above the altermagnetic N\'eel excitonic ordered state for $\Delta = 0.9$ and different polarization values. 
(a) $P_z = 1$, : altermagnetic phase;
(b) $P_z = 0.7$, $\Delta = 0.9$: altermagnetic + excitonic phase.
}
    \label{fig:flavour_waves_AM}
\end{figure*}
\noindent
{\bfseries \emph{The spectrum of the excitations} --} In order to study the dynamical response of the system, we need to incorporate quantum fluctuations around the classical ground state. To this end, we employ linear flavor wave theory (LFWT) \cite{Joshi1999,Penc2003,Miyazaki2022}.

It is therefore convenient to introduce the following change of coordinates \cite{Cherny2009,delre2016,PhysRevResearch.6.023082}, $\mathcal{U} = \prod_i \mathcal{U}_i$, constructed such that it rotates the ground state into a homogeneous configuration where all sites are in the flavor state $\ket{1}$; that is, $\mathcal{U}_i \ket{\psi_i} = \ket{1}$.
Such a unitary transformation simplifies the classical state, however it complicates the original Hamiltonian in the following way: 
\begin{align}\label{eq:Heis_rot}
\mathcal{U}\, \hat{\mathcal{P}}_{ij}\,\mathcal{U}^\dagger = \sum_{\alpha\alpha^\prime \beta \beta^\prime}\,\kappa^{\alpha \alpha^\prime}_{ij}\kappa_{ji}^{\beta\beta^\prime}S^{\alpha}_{\beta}(i)S^{\beta^\prime}_{\alpha^\prime}(j),
\end{align}
where $\kappa^{\alpha\beta}_{ij} = [\mathcal{U}_i\cdot \mathcal{U}^\dagger_j]_{\alpha\beta}$, which is a function of the difference of the positions only.

We then can express the Hamiltonian  in the rotated basis as \( E_{\text{cl}} +  H_2 \), where \(  H_2 \) captures the quantum fluctuations. These are described using a generalized Holstein-Primakoff transformation: $S^1_1 = M - \sum_n b_n^\dagger b_n$, $ S^1_m = b_m \sqrt{M - \sum_n b_n^\dagger b_n}$,  $S^n_m = b_n^\dagger b_m$, with three bosonic operators \( b_n^{(\dagger)} \), for \( n \in \{2,3,4\} \), and an expansion in powers of \( 1/M \). 
Within the harmonic approximation, the flavor-wave spectrum for different phases can be obtained by diagonalizing \(  H_2 \) as a function of \( \delta \), \( J_2 \)  and $\Delta$.
The Hamiltonian of fluctuations around the classical ground state reads:
\begin{align}\label{eq:ham_qf}
     H_2 &= \sum_{k}\xi_{-k}^T \,\mathcal{H}_k\, \xi^{\,}_{k}, 
\end{align}
where $\xi_k = (\mathbf{p}_k,\mathbf{x}_k)$. $\mathbf{x}_k$ and $\mathbf{p}_k$  are vectors whose components are given by conjugates quantum variables obeying the commutation relation $[x_{ka},p_{k^\prime b }] = i\delta_{k,-k^\prime}\delta_{ab}$. In particular:
\begin{align}
\mathbf{x}_k = (\overbrace{x_{kA2},x_{kA3},x_{kA4}}^{\text{sub-lattice A}},\overbrace{x_{kB2},x_{kB3},x_{kB4}}^{\text{sub-lattice B}})
\end{align}
is a six-dimensional vector which is made up of $3$ flavor times $2$ sub-lattice indices, where $A,B$ are the sub-lattice indices  and 2,3,4 are the flavor indices \cite{suppl}. 

We identify six possible types of excitations, corresponding to the following combinations: two intra-layer spin-flip processes, two inter-layer spin-flip processes, and two inter-layer spin-conserving (spinless) processes.
Since spin is conserved, we can label these excitations by their spin quantum number: positive or negative for spin-flip processes, and zero for spin-conserving inter-layer excitations.

We shall first consider the case at $\Delta = 0$, which is not altermagnetic. In Figure \ref{fig:flavour_waves_AF}, we show the spectrum of the spin waves for two different values of polarization. At full polarization, \( P_z = 1 \), where one of the two layers is completely empty, the excitonic order vanishes and only two excitation branches remain. The first is a doubly degenerate spinful branch, corresponding to the antiferromagnetic spin waves of the SU(2) Heisenberg model on the square lattice. The second is a fourfold-degenerate branch that captures all four possible inter-layer processes. At \( P_z = 0.7 \), the excitonic order parameter is finite and that results into three independent, doubly degenerate excitation branches: two are spinful, and one is spinless. 

For values of \(\Delta \neq 0\), altermagnetic states are allowed. In Figure \ref{fig:flavour_waves_AM}, we show the spin-wave spectra for \(P_z = 1\) and \(P_z = 0.7\) with \(\Delta = 0.9\). At \(P_z = 1\), the inter-layer excitonic coherences vanish, resulting in four branches. The intra-layer spin-flip processes are now split, reproducing the spin waves of the altermagnetic phase of the SU(2) Heisenberg model studied in Ref.~\cite{liu2024}. In this phase, the altermagnetic branches are non-degenerate and spinful, alternating across the Brillouin zone. Interestingly, the inter-layer processes--which are absent in the single-layer limit--also split into two two-fold degenerate branches that alternate across the Brillouin zone. Each branch consists of both a spinful and a spinless component, and therefore they possess an overall chirality. In particular, the inter-layer spinless processes split further, distinguishing whether a spin-up or spin-down excitation is promoted to the top layer. Since we have chosen the minimum field  $\delta$ required to achieve full polarization, the inter-layer excitations are massless. However, by increasing this external field further, an arbitrary gap can be introduced, enhancing the system's tunability and potentially making it useful for technological applications.

At $P_z = 0.7$, the system enters an altermagnetic excitonic phase, where all six excitation branches become split yet remain grouped in pairs. Notably, the spinless excitations split and alternate across the Brillouin zone, mirroring the behavior of both inter-layer and intra-layer spin excitations.
\\   \par
\noindent
{\bfseries\emph{Counter-flow experiment} --}
\begin{figure*}[htbp]
    \centering
    \includegraphics[width=0.75\columnwidth]{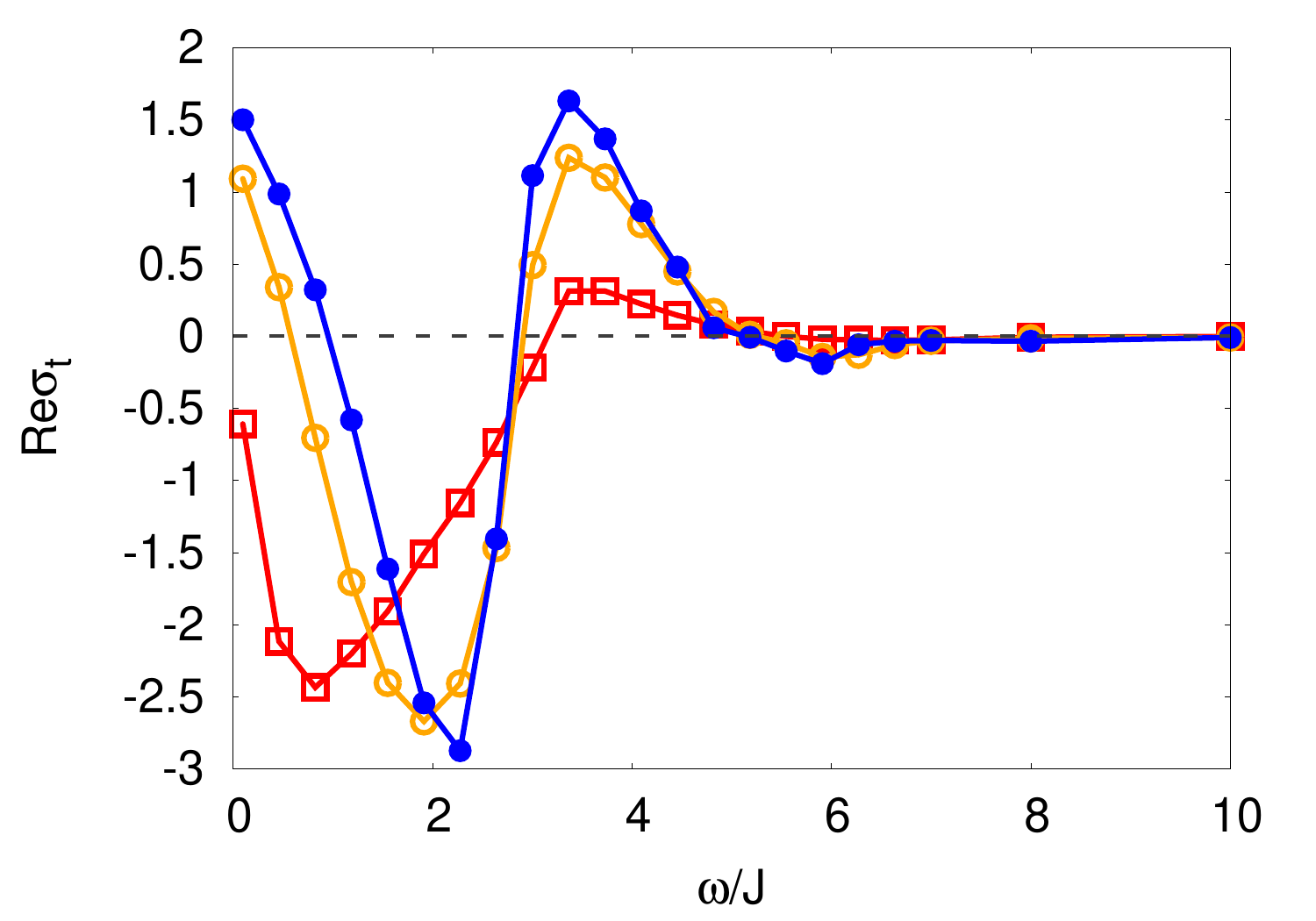}(a)
    \includegraphics[width=0.75\columnwidth]{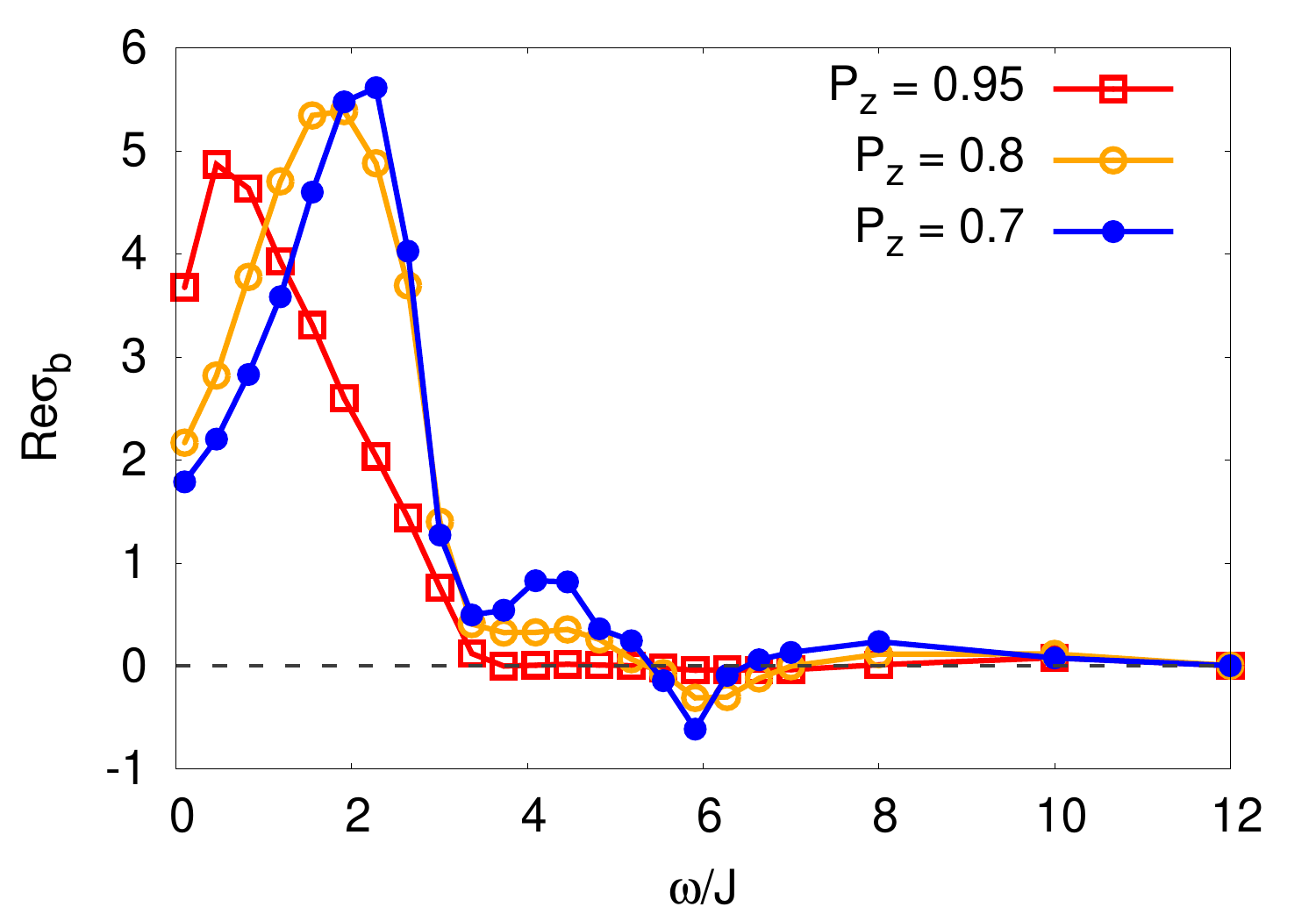}(b)
    \caption{(a)-- Real part of the spin-conductivity in the top-layer (minority species) as a function of the electric field frequency and for different values of the layer polarization.
    (b)--
    Real part of the spin-conductivity in the bottom-layer (majority species) as a function of frequency and for different values of the layer polarization.}
    \label{fig:spin_curr}
\end{figure*}
In Figure \ref{fig:sketch}, we illustrate a typical counterflow configuration in which a uniform electric field is applied to each layer in opposite directions, i.e., $\mathbf{E}_\text{t} = -\mathbf{E}_\text{b} = -\mathbf{E}$.
This setup has been previously proposed as a probe for spin-liquid phases in twisted bilayers of transition metal dichalcogenides \cite{Zhang2021}, and has been successfully implemented in quantum Hall bilayers \cite{eisenstein2014, liu2017quantum, li2017excitonic}.

Here, we propose such a counterflow setup as a means to induce a spin current--an effect that would not be possible in a single layer. In a single layer, charges are completely frozen in the Mott insulating regime, so an electric field coupling to the total charge does not induce a linear response. In contrast, in a bilayer (or multilayer) system, although the total charge remains frozen, the dipole charge--i.e., the charge associated with the polarization $P_z$ can still flow.

Let \( A_{\alpha}^\mu(t) \) denote the flavor-dependent vector potential, related to the electric field by \( E^\mu_\alpha(t) = \partial_t A^\mu_\alpha(t) \). In the counterflow configuration considered here, we impose \( A^\mu_{1} = A^\mu_{2} = A^\mu = -A^\mu_{3} = -A^\mu_{4} \). 
However, for the sake of setting up the formalism, it is useful to keep the vector potential in its most general, flavor-dependent form.

The Heisenberg Hamiltonian,  is modified by the presence of the field \cite{suppl} and assumes the following form:
\begin{align}\label{eq:Heis_out_of_eq}
    \hat{H}(A) = \sum_{ij}\sum_{\alpha\beta}J_{ij} e^{i(\mathbf{R}_i-\mathbf{R}_j)\cdot(\mathbf{A}_\alpha-\mathbf{A}_\beta)} S^{\alpha}_\beta(i)S^\beta_\alpha(j).
\end{align}
It is clear from the last equation that if $A_\alpha$
  does not depend on the flavor index -- i.e., it couples equally to the total charge of the system -- the Heisenberg Hamiltonian remains unchanged. However, in a more general configuration where the field differs between the two layers, the Heisenberg Hamiltonian is modified from its equilibrium form.
Therefore, we can define the flavor current operator, that within the harmonic approximation reads:
\begin{align}\label{eq:current_operator}
    \hat{I}^\mu_\alpha  = \sum_k \xi^T_{-k}\, I^\mu_\alpha(k)\, \xi_{k},
\end{align}
where we defined the matrix associated to the current operator as $I^\mu_\alpha(k) ={\partial_{A^\mu_\alpha}} \mathcal{H}_{k}(A)$, with $\mathcal{H}_{k}(A)$ being the Hamiltonian matrix of the quantum fluctuations modified by the presence of the external field.

Following Kubo formalism, the regular part of the flavor resolved conductivity reads \cite{suppl}:
\begin{align}\label{eq:conductivity}
    \text{Re}\,\sigma^{\mu\nu}_{\alpha\beta,\text{reg.}}(\omega) = \frac{\text{Im}\,\Pi^{\mu\nu}_{\alpha\beta}(\omega)}{\omega} ,  
\end{align}
where  $\Pi^{\mu\nu}_{\alpha\beta}(\omega)$ is the Fourier transform of the flavor-resolved current-current correlation function, i.e. $\Pi^{\mu\nu}_{\alpha\beta}(t) = -i\theta(t)\left<[\hat{I}^{\mu}_\alpha(t),\hat{I}^{\nu}_\beta(0)]\right>$.

Within the harmonic approximation, we can obtain an analytic expression for the current-current correlation function that reads:
\begin{align}
    \Pi^{\mu\nu}_{\alpha\beta}(\omega) =2\sum_{k}\sum_{ab}^\prime&\frac{\left[\overline{I}^{\alpha}_{\mu}(k)\right]_{ab}\left[\overline{I}^{\beta}_{\nu}(k)\right]_{ba} }{\epsilon_{ka} + \epsilon_{kb} + \omega + i\eta}\nonumber \,+\\  &
    \frac{\left[\overline{I}^{\alpha}_{\mu}(k)\right]^*_{ab}\left[\overline{I}^{\beta}_{\nu}(k)\right]_{ba}^* }{\epsilon_{ka} + \epsilon_{kb} - \omega -i\eta} ,
\end{align}
where, $\overline{I}^\mu_\alpha(k) = T^\dagger S^T_k I^\mu_{\alpha}(k)S_kT$ is the expression of the current in the quasi-particle basis. The prime in the sum $\sum_{ab}^\prime$ , and the details about the transformation of the current operators are discussed in the supplemental materials \cite{suppl}.

Since we are interested in the possibility of inducing a spin current as a consequence of a polarization flow, we consider the following contraction of indices of the flavor-dependent conductivity tensor:
\begin{equation}\label{eq:spin_conductivity}
\sigma^{\mu\nu}_\ell(\omega) = \sum_{\alpha\beta} P^z_\alpha S^\ell_\beta\, \sigma^{\mu\nu}_{\alpha\beta}(\omega),
\end{equation}
where $\ell = t, b$ labels the top and bottom layers, respectively. The coefficients are defined as
\[
P^z_\alpha = \{1,1,-1,-1\}, \ S^t_\alpha = \{1,-1,0,0\}, \ S^b_\alpha = \{0,0,1,-1\}.
\]

In essence, $\mathrm{Re}\,\sigma^{\mu\nu}_\ell(\omega)$ quantifies how an electric field can generate a spin current in the $\ell$-th layer. We therefore refer to it as the {layer-resolved spin conductivity}. The sign of this quantity is also significant, as it determines the direction of the resulting spin current.


In Figure \ref{fig:spin_curr}, we show the spin conductivity as a function of photon energy for the top (bottom) layer, corresponding to the minority (majority) species.

As pairs of orthogonal axes we chose the second-neighbor directions $\mu = \tilde{x},\tilde{y}$ where $
\tilde{x} = \frac{1}{\sqrt{2}}(1,1), \quad \tilde{y} = \frac{1}{\sqrt{2}}(1,-1)
$. In this basis the conductivity tensor is purely diagonal but with opposite diagonal entries, i.e. $\sigma_\ell^{\tilde{x}\tilde{x}} = -\sigma_\ell^{\tilde{y}\tilde{y}} = -\sigma_\ell$.


We observe that the spin current depends strongly on the field frequency for both species. In particular, for the minority species, the field frequency can induce a reversal of the spin-current sign, especially for moderate values of the polarization. Interestingly, the spin current for the minority species can be switched off while remaining finite for the majority species. Additionally, the spin-current maxima (minima) for the majority (minority) species can be tuned by adjusting the photon energy and shift toward higher energies as the polarization decreases.
\\ \par 
\noindent
{\bfseries \emph{Conclusions} --} We have studied a four-flavor Heisenberg model with an on-site energy that distinguishes pairs of flavors, interpreted as spins in different layers. Introducing an anisotropic second-neighbor exchange term that breaks translational symmetry, we stabilize altermagnetic phases. Using mean-field theory and linear flavor-wave theory, we show that the system supports an altermagnetic state with antiferromagnetic ordering of spins in both layers.

The physics is further enriched by the condensation of inter-layer excitons, identified as coherent inter-layer processes arising from spontaneous symmetry breaking. The excitation spectrum exhibits unique features combining both excitonic and altermagnetic characteristics.

Finally, we propose a mechanism to harness this rich physics by generating a spin current in response to a counterflow electric field--a hallmark of bi-layer altermagnetic Mott insulators, absent in single-layer or conventional antiferromagnetic insulators. Our findings open new avenues for exploiting altermagnetic and excitonic phases in spintronic applications.
\par  
\noindent
{\bfseries \emph{Acknowledgment} --} We thank Alessandro Toschi for valuable discussions. This work is funded by the Deutsche Forschungsgemeinschaft (DFG, German Research Foundation) through SFB 1170 ToCoTronics, the W\"urzburg-Dresden Cluster of Excellence on Complexity and Topology in Quantum Matter--ct.qmat Project-ID 390858490--EXC 2147.

\bibliography{ref_biblio.bib}
\end{document}


\title{Optically Tunable Spin Transport in Bilayer Altermagnetic Mott Insulators\\
{\it -- Supplemental Material --}}

\author{Niklas Sicheler, Roberto Raimondi,  Giorgio Sangiovanni and Lorenzo~Del~Re}
\maketitle

\section{Linear flavor wave theory}


In the following, we present a detailed derivation of the magnon excitation spectrum.

As a first step, we rotate the ground state such that all sites are in the flavor state $\ket{1}$ after the transformation. This simplifies the application of the linear flavor wave theory.
To perform this rotation, we apply a unitary transformtion  $\mathcal{U}=\prod_i \mathcal{U}_i$, where each $\mathcal{U}_i$ maps the local Néel excitonic state $\ket{\psi_i}$ on site $i$ to $\ket{1}$, i.e. $\mathcal{U}_i\ket{\psi_i}=\ket{1}$.
Its Hermitian conjugate is given by $
    \mathcal{U}_i^\dagger = \mathcal{Q} \mathcal{A}(\mathbf{R}_i) \mathcal{R}$, where 
\begin{equation}
    \mathcal{A}(\mathbf{R_i}) = \begin{pmatrix}
        1 & & & \\
        & e^{iQ_s^t R_i} & & \\
        & & e^{iQ_pR_i} & \\
        & & & e^{i(Q_s^b+Q_p)R_i}
    \end{pmatrix},
    \label{eq:A-matrix}
\end{equation}
$
    \mathcal{R} = \mathcal{R}^{o}_y\left(\vartheta\right) \otimes \mathcal{R}^{s}_y\left(\frac{\pi}{2}\right) 
$
and 
$ \mathcal{Q}=\mathbb{I}_{2\times2}\otimes \mathcal{R}^s_y(-\frac{\pi}{2})$.
Here, $\vartheta$ has to be chosen as
$\vartheta=2\arccos\sqrt{\frac{1-P_z}{2}}$ and $\mathcal{R}_y^{s/o}(\vartheta)$ represents a spin/pseudospin rotation of $\vartheta$ around the $y$-axis.
It is easy to see that the product $\mathcal{U}_i\mathcal{U}_j^\dagger$ is only dependent on the difference between the two lattice sites:
$
 \mathcal{U}_i \mathcal{U}_j^\dagger= \mathcal{R}^\dagger \mathcal{A}(\mathbf{R}_j-\mathbf{R}_i)\mathcal{R}.
$

We start from the Hamiltonian 
$\hat{H}=\sum_{ij} J_{ij} \hat{\mathcal{P}}_{ij}+\delta\sum_i\hat{P}_i^z$, where $\hat{\mathcal{P}}_{ij}=S^\alpha_\beta(i)S^\beta_\alpha(j)$ with $S_\beta^\alpha=\ket{\alpha}\bra{\beta}$ and $\hat{P}_i^z=\ket{1}\bra{1}+\ket{2}\bra{2}-\ket{3}\bra{3}-\ket{4}\bra{4}$. Applying the unitary transformation leads to:
\begin{align}
    \mathcal{U}H\mathcal{U}^\dagger&=\sum_{ij} \sum_{\alpha\alpha^\prime \beta \beta^\prime}\,\kappa^{\alpha \alpha^\prime}_{ij}\kappa_{ji}^{\beta\beta^\prime}S^{\alpha}_{\beta}(i)S^{\beta^\prime}_{\alpha^\prime}(j) \nonumber\\ & +\delta \sum_i \sum_{\alpha \beta} \tilde{P}^z_\ab S^\alpha_\beta(i)
    \label{eq:Hamiltonian_trans},
\end{align}
where $\kappa^{\alpha\beta}_{ij}=\braket{\alpha|\mathcal{U}_i \mathcal{U}_j^\dagger|\beta}$ and $\tilde{P}_\ab^z=\braket{\alpha|\mathcal{U}_i \hat{P}_i^z \mathcal{U}_i^\dagger|\beta}$. 
In the following calculation, we use the easy to show properties $ {\left(\kappa^\ab_{ij}\right)^*=\kappa^\ba_{ji}}$, $ {\left(\tilde{P}^z_\ab\right)^* = \tilde{P}^z_\ba}$ and the fact that $\kappa^{\alpha\beta}_{ij}$ only depends on the difference $\mathbf{R}_j-\mathbf{R}_i$.

Now, we perform a generalized Holstein-Primakoff transformation \cite{joshi1999,Penc2003}, which is given by:
\begin{align}
    S^1_1(i)&=M-\sum_{\alpha \neq 1} b^\dagger_{i\alpha} b_{i\alpha} \nonumber,\\
    S^1_\alpha(i) &= \sqrt{M} b_{i\alpha} \sqrt{1-{\sum\nolimits_{\alpha}} b^\dagger_{i\alpha} b_{i\alpha}} \sim b_{i\alpha} \nonumber \enspace \text{for } \alpha\neq1,\\
    S^\alpha_\beta(i) &= b^\dagger_{i\alpha}b_{i\beta} \enspace \text{for } \alpha,\beta\neq1,
    \label{eq:holstein-trans}
\end{align}
where $M$ is the classical expectation value of $S^1_1$ (in our case $M=1$).
$b_{i\alpha}$ and $b_{i\alpha}^\dagger$ are the bosonic operators that create/destroy a flavor deviation on site $i$. They obey the commutation relation $\left[b_{i\alpha},b^\dagger_{j\beta}\right]=\delta_{i,j}\delta_{\alpha,\beta}$. 

By substituting Eq.\eqref{eq:holstein-trans} into Eq.\eqref{eq:Hamiltonian_trans}, we can expand $\mathcal{U}H\mathcal{U}^\dagger$ into a series:
$\mathcal{U}H\mathcal{U}^\dagger=H_0+H_1+H_2+\dots$. In this series, $H_n$ contains bosonic operators to the power of $n$. We use the harmonic approximation and therefore only develop up to the second order.
After setting $M=1$, we obtain for the zeroth order:
\begin{align}
    H_0 &= \sum_{ij} J_{ij} \kappa^{11}_{ij} \kappa^{11}_{ji}+ \delta \sum_i \tilde{P}^z_{11}\nonumber\\&= N\sum_{\tau} J(\tau) \abs{\kappa_{11}(\tau)}^2+\delta N \tilde{P}^z_{11},
\end{align}
where $\sum_\tau$ is the sum over the nearest and next-nearest neighbors with $\tau=R_j-R_i$. This is exactly the classical energy.
The first and second order read:
\begin{align}
    H_1 
    &= 2 \sum_{ij} J_{ij} \sum_{\alpha \neq 1} \left(\kappa^{11}_{ij} {(\kappa^{\alpha1}_{ij})}^* b_{i\alpha} + \text{H.c.} \right) \nonumber\\&+ \delta \sum_i \sum_{\alpha\neq1} \left(\tilde{P}^z_{1\alpha} b_{i\alpha} + \text{H.c.}\right)
\end{align}
and
\begin{align}
    H_2 &= \sum_{ij} J_{ij} \nonumber\\ &\times \left\{ -2 \abs{\kappa^{11}_{ij}}^2 \sum_\alpha b_{i\alpha}^\dagger b_{i\alpha}\nonumber\right.+ 2\sum_\ab \kappa^{\alpha1}_{ij} {(\kappa^{\beta1}_{ij})}^* b_{i\alpha}^\dagger b_{i\beta}  \nonumber\\ &+ \sum_\ab \left[{(\kappa^{\alpha1}_{ij})}^* \kappa^{1\beta}_{ij} b_{i\alpha} b_{j\beta} + \kappa^{\alpha1}_{ij} {(\kappa^{1\beta}_{ij})}^* b^\dagger_{i\alpha} b^\dagger_{j\beta}\right] \nonumber\\&
     + \left. \sum_\ab \left[ \xi^{11}_{ij} {(\kappa^\ab_{ij})}^* b_{i\alpha} b^\dagger_{j\beta} + {(\kappa^{11}_{ij})}^* \kappa^\ab_{ij} b^\dagger_{i\alpha} b_{j\beta}\right]
    \right\} \nonumber\\&+ \delta \sum_i \left[ -\sum_\alpha \tilde{P}^z_{11} b_{i\alpha}^\dagger b_{i\alpha} + \sum_\ab \tilde{P}^z_\ab   b_{i\alpha}^\dagger b_{i\beta} \right]
    \label{eq:H_second_order}
\end{align}
with $\alpha,\beta\neq1.$

A non-vanishing $H_1$ would lead to an unstable system, therefore this leads to the condition:
\begin{equation}
    \delta=-\frac{\sum_{\tau} 2 J(\tau) \, \mathrm{Re} (\kappa_{11}(\tau)\kappa_{\alpha1}^*(\tau)) }{\tilde{P}^z_{1\alpha}}.
    \label{eq:delta-determination}
\end{equation} The condition must be fulfilled for $\alpha=2,3,4$.

Before we can apply a Fourier transform, we explictly introduce a sublattice index $\lambda\in\{A,B\}$. Each operator is labeled by its site $i$, the sublattice $\lambda$ and the flavor $\alpha$. The sums in $H_2$ in Eq.\eqref{eq:H_second_order} therefore split into four contributions $AA$, $AB$, $BA$ and $BB$, depending wether the operators act within the same or a different sublattice.

Replacing the bosonic operators in Eq.\eqref{eq:H_second_order} with their Fourier series $ b_{iA\alpha}=\frac{1}{\sqrt{N}}\sum_{k} e^{ikR_i}b_{kA\alpha} ,\,\,b^\dagger_{iA\alpha}=\frac{1}{\sqrt{N}}\sum_{k} e^{-ikR_i}b^\dagger_{kA\alpha}
$ 
and using the identity $ \sum_i e^{i(k-k')R_i}=N\delta_{k,k'} $ yields:

\begin{align}
    H_2 =& \sum_{k\ab} \left(f^{A}_\ab+u^{A}_\ab\right) b_{kA\alpha}^\dagger b_{kA\beta} \nonumber\\ +& \sum_{k\ab} \left(h^{A}_\ab(k)b_{kA\alpha}^\dagger b_{kA\beta}+g^{A}_\ab(k) b_{kA\alpha} b_{-kA\beta} \right. \nonumber\\
    & \left.+ \, w^{A}_\ab(k) b_{kA\alpha}^\dagger b_{kB\beta} + v^{A}_\ab(k) b_{kA\alpha} b_{-kB\beta} +\text{H.c.}\right) \nonumber\\
     + &\,  \text{terms with B$\leftrightarrow$A},
    \label{eq:Fourier-hamiltonian}
\end{align}
where
\begin{align}
    f^{A/B}_\ab &= \sum_{\rho} 2 J^{AA/BB}(\rho) \nonumber\\& \times \left(- \delta_{\alpha,\beta} \abs{\kappa^{A/B}_{11}(\rho)}^2+ \kappa^{A/B}_{\alpha1}(\rho) \left[{\kappa^{A/B}_{\beta1}}(\rho)\right]^*  \right)\nonumber \\&+ \delta \left(-\delta_{\alpha,\beta}\tilde{P}^z_{11} + \tilde{P}^z_\ab \right), \nonumber\\
    g^{A/B}_\ab(k) &= \sum_{\rho} J^{AA/BB}(\rho) \left[{\kappa^{A/B}_{\alpha1}}(\rho)\right]^*\kappa^{A/B}_{1\beta}(\rho) e^{-ik\rho} ,\nonumber \\
    h^{A/B}_\ab(k) &= \sum_{\rho} J^{AA/BB}(\rho) \left[{\kappa^{A/B}_{11}}(\rho)\right]^* \kappa^{A/B}_\ab(\rho) e^{ik\rho}\nonumber,\\
    u^{A/B}_\ab& =\sum_{\tau} 2J_1   \nonumber \\ & \times  \left( -\delta_{\alpha,\beta} \abs{\kappa^{A/B}_{11}(\tau)}^2+ \kappa^{A/B}_{\alpha1}(\tau)\left[{\kappa^{A/B}_{\beta1}}(\tau)\right]^* \right), \nonumber\\
    v^{A/B}_\ab(k) &= J_1 \sum_{\tau} \left[{\kappa^{A/B}_{\alpha1}}(\tau)\right]^*\kappa^{A/B}_{1\beta}(\tau) e^{-ik\tau}, \nonumber\\
    w^{A/B}_\ab(k) &= J_1 \sum_{\tau}\left[{\kappa^{A/B}_{11}}(\tau)\right]^* \kappa^{A/B}_\ab(\tau) e^{ik\tau}.
\end{align}
Here, $\sum_\rho$ denotes the sum over the next-nearest neighbors and $\sum_\tau$ is the sum over the nearest neighbors. Additionally, we define 
$J^{AA}=J_2(1+\Delta),J^{BB}=J_2(1-\Delta)$ for $\rho=\pm d_1$ and $J^{AA}=J_2(1-\Delta),J^{BB}=J_2(1+\Delta)$ for $\rho= \pm d_2$ with $d_1=(1,1)$ and $d_2=(1,-1)$. At this point, the $\kappa$'s are not dependent on the sublattice because they are only a function of the difference $\mathbf{R}_j-\mathbf{R}_i$. However, once we introduce the vector potential later on, they become sublattice dependent. 

It is useful to express the bosonic operators in terms of the canonical position and momentum operators: $
b_{kA\alpha}=\frac{1}{\sqrt{2}}\left(x_{kA\alpha}+i p_{kA\alpha}\right),\,\,b^\dagger_{kA\alpha}=\frac{1}{\sqrt{2}}\left(x_{-kA\alpha}-i p_{-kA\alpha}\right)$.
The new operators obey the canonical commutation relations: $\left[x_{kA\alpha},p_{k'B\beta}\right]=i \delta_{k,-k'}\delta_{A,B}\delta_{\alpha,\beta}$.
The Hamiltonian in the new basis can be written in a more compact way as a $12\times12$ matrix. For this, we introduce a 12 dimensional vector $\xi_k = (\mathbf{p}_k,\mathbf{x}_k)$, where $\mathbf{p}_{k}=(p_{kA2},p_{kA3},p_{kA4},p_{kB2},p_{kB3},p_{kB4})$ and $\mathbf{x}_k$ is defined in the same way. $A$ and $B$ are the sublattice indices and 2, 3, 4 are the flavor indices.
Then, we can write:
\begin{align}\label{eq:ham_qf}
     H_2 &= \sum_{k} \xi^T_{-k} \,\mathcal{H}_k\,\xi_k  +C, 
\end{align}
where $\mathcal{H}_k$ is a $12\times12$ matrix and $C$ a constant with value $C=-\frac{N}{4}\sum_{\alpha\lambda} (f^\lambda_{\alpha\alpha}+u^\lambda_{\alpha\alpha})$.
The block matrix $\mathcal{H}_k$ has the structure:
\begin{align}
\begin{pmatrix}
    H^{PP} & H^{PX} \\
    H^{XP} & H^{XX}
\end{pmatrix},
\end{align}
where each block is a block matrix itself that consists of four $3\times3$ matrices:
\begin{align}
    H^{YZ} = \begin{pmatrix}
        H_{AA}^{YZ} & H^{YZ}_{AB} \\
        H^{YZ}_{BA} & H^{YZ}_{BB}
    \end{pmatrix} .
\end{align}

The matrix elements are given by:
\begin{align}
    {\left(H_{AA/BB}^{PP}\right)}_\ab =& f^{A/B}_\ab + u^{A/B}_\ab + h^{A/B}_\ab(k)+h^{A/B}_\ab(-k)\nonumber\\&-g^{A/B}_\ab(-k)-g^{A/B}_\ab(k),  \nonumber\\
    {\left(H_{AB/BA}^{PP}\right)}_\ab  =&  w^{A/B}_\ab(k)+w^{A/B}_\ab(-k)\nonumber\\&-v^{A/B}_\ab(k) - v^{A/B}_\ab(-k),\nonumber\\
    {\left(H_{AA/BB}^{XX}\right)}_\ab =& f^{A/B}_\ab + u^{A/B}_\ab + h^{A/B}_\ab(k)+h^{A/B}_\ab(-k)\nonumber\\&+g^{A/B}_\ab(-k)+g^{A/B}_\ab(k)  ,\nonumber\\
    {\left(H_{AB/BA}^{XX}\right)}_\ab =& w^{A/B}_\ab(k) + w^{A/B}_\ab(-k)\nonumber\\&+v^{A/B}_\ab(-k) + v^{A/B}_\ab(k) ,\nonumber\\
    {\left(H_{AA/BB}^{PX}\right)}_\ab  =&-i\left[ h^{A/B}_\ab(k)-h^{A/B}_\ab(-k) \right. \nonumber\\ &\left.+g^{A/B}_\ab(k)-g^{A/B}_\ab(-k)  \right],\nonumber\\
    {\left(H_{AB/BA}^{PX}\right)}_\ab =& -i\left[w^{A/B}_\ab(k) - w^{A/B}_\ab(-k) \right.\nonumber\\&\left. + v^{A/B}_\ab(k)- v^{A/B}_\ab(-k)\right] ,\nonumber\\
    {\left(H_{AA/BB}^{XP}\right)}_\ab  =&\left[{\left(H_{AA/BB}^{PX}\right)}_\ba\right]^*,\nonumber\\
    {\left(H_{AB/BA}^{XP}\right)}_\ab=& \left[{\left(H_{BA/AB}^{PX}\right)}_\ba\right]^* .
\end{align}
Here, we used the fact that $f$ and $u$ are real and $k$-independent matrices, and that $h$, $g$, $w$, and $v$ are real matrices.


\section{Linear response theory for the counter-flow experiment}
\subsection{Current operator}
Here, we shall derive in more detail the expression of the flavor-resolved current-current correlation presented in the main text. 

Let us recall that in a single layer, charges are completely frozen in the Mott insulating regime, so an electric field coupling to the total charge does not induce a linear response. In contrast, in a bilayer (or multilayer) system, although the total charge remains frozen, the dipole charge--i.e., the charge associated with the polarization $P_z$ can still flow.

We can better illustrate this idea by starting from the SU(4)-Hubbard model in presence of the counter-flow electric field, that reads:

\begin{align}\label{eq:SU(4)_Hub}
    \hat{H}_\text{Hubb} = \sum_{ij}\sum_\alpha t_{ij}e^{i\mathbf{A}_\alpha(t)(\mathbf{R}_i-\mathbf{R}_j)}c^\dagger_{i\alpha}c^{\,}_{j\alpha} + H_\text{int},
\end{align}
where $A_{\alpha}(t)$ is the flavor-dependent vector potential, which is related to the field via $E^\mu_\alpha = \partial_t A^\mu_\alpha(t)$ and in particular, in our counter-flow scheme we have that $A^\mu_{1} = A^\mu_2 = A^\mu =-A^\mu_3 =-A^\mu_4$.

Hence, the Heisenberg Hamiltonian, which is obtained via second-order perturbation theory starting from Eq.(\ref{eq:SU(4)_Hub}), is modified by the presence of the field and assumes the following form:

\begin{align}\label{eq:Heis_out_of_eq}
    \hat{H}(A) =P_0 \sum_{ij}\sum_{\alpha\beta}\frac{t_{ij}t_{ji}}{U} e^{i(\mathbf{R}_i-\mathbf{R}_j)\cdot(\mathbf{A}_\alpha-\mathbf{A}_\beta)} S^{\alpha}_\beta(i)S^\beta_\alpha(j)\, P_0,
\end{align}
where $P_0$ is the projector onto the singly occupied sites manifold.
It is clear from the last equation that if $A_\alpha$
  does not depend on the flavor index -- i.e., it couples equally to the total charge of the system -- the Heisenberg Hamiltonian remains unchanged. However, in a more general configuration where the field differs between the two layers, the Heisenberg Hamiltonian is modified from its equilibrium form.

It is easy to show that the out-of-equilibrium Hamiltonian in the new basis has the same form as the one at equilibrium, provided that we modify the $\kappa$ matrices in the following way: $\kappa_{ij}(A^\mu_\alpha) = \mathcal{U}_i\exp[i\sum_\alpha P_\alpha  A^\mu_\alpha (R_i^\mu-R_j^\mu) ]\mathcal{U}^\dagger_j $, where we introduced the following $4\times 4$ projector matrix $[P_{\gamma}]_{\alpha\beta} = \delta_{\alpha\beta}\delta_{\alpha\gamma}$, where we left the flavor dependence of the vector potential generic. Hence, the Hamiltonian of the quantum fluctuations $H_2$ in presence of an external field has the same form as the one in Eq.(\ref{eq:ham_qf}) provided we modify $\mathcal{H}_k(A^\mu_\alpha)$ using the new expression for $\kappa$. 
If we  expand the Hamiltonian up to second order in $A^\mu_\alpha$ we obtain:
\begin{align}
    H(A)&- H(A = 0)\sim  \nonumber \\ &  \sum_{\alpha\mu} A^\mu_\alpha\left(\sum_k \xi^T_{-k}\, I^\mu_\alpha(k)\, \xi_{k} +
    \sum_{\beta\nu} A^\nu_\beta\sum_k \xi^T_{-k} \,\mathcal{I}^{\mu\nu}_{\alpha\beta}\xi_k\right).
\end{align}
In the last equation, we recognize the current operator as the vector coupling to the vector potential, and it has two components: the first is the paramagnetic current, that reads:
\begin{align}\label{eq:current_operator}
    \hat{I}^\mu_\alpha  = \sum_k \xi^T_{-k}\, I^\mu_\alpha(k)\, \xi_{k},
\end{align}
where $I^\mu_\alpha(k) ={\partial_{A^\mu_\alpha}} \mathcal{H}_{k}(A)$;
the second is the diamagnetic current which is given by:
\begin{align}
    \hat{\mathcal{I}}^\mu_{\alpha} =  \sum_{\beta\nu} A^\nu_\beta\sum_k \xi^T_{-k} \,\mathcal{I}^{\mu\nu}_{\alpha\beta}(k)\,\xi_k,
\end{align}
where $\mathcal{I}^{\mu\nu}_{\alpha\beta}(k) = \frac{1}{2}\frac{\partial^2\mathcal{H}_k(A)}{\partial A^\mu_\alpha\partial A^\nu_\beta}$.

It is possible to express the current operators as derivatives with respect to the crystalline momenta. In fact, by computing the derivatives of the non-equlibrium Hamiltonian in real space we have:
$
    \frac{\partial \mathcal{H}_{ij}(A)}{\partial A^\mu_\alpha} = \sum_{\beta\gamma}\frac{\partial\mathcal{H}_{ij}}{\partial\kappa^{\beta\gamma}_{ij}}\frac{\partial \kappa_{ij}^{\beta\gamma}}{\partial A^\mu_\alpha}
$, and since $\partial_{A^\mu_\alpha}\kappa^{\beta\gamma}_{ij} = i(R_i-R_j)^\mu \,U^{\beta\alpha}_{i}[U^{\dagger}]^{\alpha\gamma}$. Hence, we can express the paramagnetic current operator as:

\begin{align}
    \hat{I}^\mu_\alpha = \sum_{ij}i(R_i-R_j)^\mu \xi^T_i\,F^\alpha_{ij}\,\xi_j,
\end{align}
where:
\begin{align}
    F^{\alpha}_{ij}  = \sum_{\beta\gamma}U^{\beta\alpha}_{i}[U^{\dagger}_j]^{\alpha\gamma} \frac{\partial \mathcal{H}_{ij}}{\partial \kappa_{ij}^{\beta\gamma}}\,.
\end{align}
Finally, we can rewrite the expression for the paramagnetic current in Fourier space as:
\begin{align}
    \hat{I}^\mu_\alpha =    \sum_k\xi^T_{-k}\,\frac{\partial F_k^\alpha}{\partial k^\mu}\,\xi_{k}, 
\end{align}
where $F^\alpha_k$ is the Fourier transform of $F^\alpha_{ij}$ and  hence we can identify the following expression for the current matrix representation: $I^\mu_\alpha(k) = \frac{\partial F_k^\alpha}{\partial k ^\mu}$. 
\subsection{Conductivity}
Following Kubo formalism, we can express the time dependent average of the paramagnetic current as following:

\begin{align}
    \left<\hat{I}^\mu_\alpha(t) \right> = \sum_{\nu\beta}\int_{-\infty}^{+\infty}dt^\prime \Pi^{\mu\nu}_{\alpha\beta}(t-t^\prime)A^\nu_\beta(t^\prime),
\end{align}
where we defined the current-current correlation function:
\begin{align}
    \Pi^{\mu\nu}_{\alpha\beta}(t-t^\prime) = -i\theta(t-t^\prime)\left<[\hat{I}^\mu_\alpha(t),\hat{I}^\nu_\beta(t^\prime)]\right>.
\end{align}
The expression for the average of the time dependent diamagnetic current is more straightforward since the vector potential appears explicitly in its definition and we have:

\begin{align}
    \left<\hat{\mathcal{I}}^\mu_\alpha(t) \right> =\sum_{\nu\beta}A^\nu_\beta(t)K^{\mu\nu}_{\alpha\beta},
\end{align}
where:
\begin{align}
  K^{\mu\nu}_{\alpha\beta}  =  \left<\sum_k \xi^T_{-k}\, \mathcal{I}^{\mu\nu}_{\alpha\beta}\,\xi_{k}\right>.  
\end{align}
By expanding the time dependent averages in their Fourier components, and using $A^\mu_\alpha(\omega) =E^\mu_\alpha(\omega)/i\omega$, we obtain the following expression for the total current:
\begin{align}
    I^{\mu}_\alpha(\omega) + \mathcal{I}^{\mu}_\alpha(\omega) = \sum_{\nu\beta} \frac{\Pi^{\mu\nu}_{\alpha\beta}(\omega) + K^{\mu\nu}_{\alpha\beta}}{i\omega}E^\nu_\beta(\omega),
\end{align}
where we identify the flavor resolved conductivity that reads:
\begin{align}\label{eq:conductivity}
    \sigma^{\mu\nu}_{\alpha\beta}(\omega) = \frac{\Pi^{\mu\nu}_{\alpha\beta}(\omega) +K^{\mu\nu}_{\alpha\beta}}{i\omega} .
\end{align}

Given the expression in Eq.(\ref{eq:conductivity}), we can isolate the regular part from the singular one in the real part of the conductivity, i.e. 
\begin{align}
    \text{Re}\sigma^{\mu\nu}_{\alpha\beta}(\omega) = \pi D^{\mu\nu}_{\alpha\beta}\,\delta(\omega) +\sigma^{\mu\nu}_{\alpha\beta,\text{reg}}(\omega),
\end{align}
where $D^{\mu\nu}_{\alpha\beta} = -K^{\mu\nu}_{\alpha\beta} - \text{Re}\Pi^{\mu\nu}_{\alpha\beta}(0)$ and $\sigma^{\mu\nu}_{\alpha\beta,\text{reg}}(\omega) = \text{Im}\Pi^{\mu\nu}_{\alpha\beta}(\omega)/\omega$. 
\subsection{Current-current correlation function}

In practice it is convenient to introduce the current-current correlation function in imaginary time as following:
\begin{align}\label{eq:curr_im_time}
    \Pi^{\mu\nu}_{\alpha\beta}(\tau) =-T_\tau\left<\hat{I}^{\mu}_\alpha(\tau)\hat{I}^\nu_\beta(0)\right>,
\end{align}
then, evaluate the Fourier transform of  Eq.(\ref{eq:curr_im_time}), that reads: 
\begin{align}\label{eq:GF_FT}
    \Pi^{\mu\nu}_{\alpha\beta}(i\omega_n) = \int_0^\beta d\tau\, \Pi^{\mu\nu}_{\alpha\beta}(\tau) e^{i\omega_n\tau},
\end{align}
with  $\omega_n = 2\pi n T $ being the Matsubara frequency.
Eventually we will perform the analytic continuation i.e. $i\omega_n \to \omega + i\eta$.

By substituting Eq.(\ref{eq:current_operator}) into Eq.(\ref{eq:curr_im_time}), the correlation function in imaginary time reads:
\begin{align}
    \Pi^{\mu\nu}_{\alpha\beta}(\tau) &=-\sum_{kk^\prime}\sum_{ab}\sum_{a^\prime b^\prime}[I^{\mu}_\alpha(k)]_{ab}[I^{\nu}_\beta(k^\prime)]_{a^\prime b^\prime}  \nonumber \\ 
    &\times  T_\tau\left<\xi_{ka}(\tau)\xi_{-kb}(\tau)\xi_{k^\prime a^\prime}\xi_{-k^\prime b^\prime}\right>.
\end{align}
Using Wick's theorem, the Fourier transform in Eq.(\ref{eq:GF_FT}) we have:
\begin{align}\label{eq:Pi_omega}
    \Pi^{\mu\nu}_{\alpha\beta}(i\omega_n) &= -2\sum_{k}\sum_{a\, b\, a^\prime b^\prime}[I^\mu_\alpha(k)]_{ab}[I^\nu_\beta(k)]_{a^\prime b^\prime} \nonumber \\
    & \times \frac{1}{\beta}\sum_{m = -\infty}^{+\infty} G^{ab^\prime}_{k,m} G^{a^\prime b}_{k,m-n},
\end{align}
where we used $I^\mu_\alpha(-k) = -I_{\alpha}^\mu(k)$,  $[I^{\mu}_{\alpha}(k)]^T = -I^{\mu}_{\alpha}(k)$, since it is an hermitian and imaginary matrix and we introduced the Green's function $G^{ab}_k(\tau) = T_\tau\left<\xi_{ka}(\tau)\xi_{-kb}(0)\right>$ and its Fourier transform, $G^{ab}_{k,m} = \int_0^\beta d\tau\,e^{i\omega_m\tau}G^{ab}_{k}(\tau) $.

 We can  express {$G^{ab}_k(\tau)$} in terms of the Green's function of the quasi-particles by means of a canonical transformation that diagonalizes the Hamiltonian in Eq.(\ref{eq:ham_qf}) and that preserves the canonical variables commutation relations. {This is done via} 
a transformation $S_k$ so that $S^T_k \mathcal{J} S_k = \mathcal{J}$ and $S^T \mathcal{H}_k S_k = \text{diag}(\boldsymbol{\epsilon}_k,\boldsymbol{\epsilon}_k)$, with $\boldsymbol{\epsilon}_k$ being six dimensional and containing the quasi-particle spectrum that we have shown in the previous section.
Such a transformation defines a set of new canonical coordinates, namely $\xi_k=\tilde{\xi}_{k}\, S^T_k$ and $\xi_{-k}=S_k\,\tilde{\xi}_{-k}$. 

Hence, the Green's function \cite{PhysRevResearch.6.023082} can be rewritten in the following way:
\begin{align}\label{eq:gf1}
    G^{ab}_{k,m} &= \sum_{a_1 b_1} [S^T]_{a_1 a}[S_k]_{b b_1} \tilde{G}^{a_1 b_1}_{k,m},
\end{align}
where 
\begin{align}
\widetilde{G}^{ab}_{k,m} &= \frac{A_{ab}}{i\omega_m -\epsilon_{ka}} + \frac{B_{ab}}{i\omega_m +\epsilon_{ka}},
\end{align}

and we defined the following simple matrices:

 \begin{align}
     A = \frac{1}{2}\left(
     \begin{array}{cc}
     \mathbb{I}_{6\times 6}&-i\,\mathbb{I}_{6\times 6}\\
     i\, \mathbb{I}_{6\times 6}&\mathbb{I}_{6\times 6}
     \end{array}
     \right), \\
      B = \frac{1}{2}\left(
     \begin{array}{cc}
     -\mathbb{I}_{6\times 6}&-i\,\mathbb{I}_{6\times 6}\\
     i\, \mathbb{I}_{6\times 6}&-\mathbb{I}_{6\times 6}
     \end{array}
     \right).
 \end{align}

 As a last step, we can easily simultaneously diagonalize the matrices $A (B) = \frac{1}{2}(\pm\mathbb{I}_{2\times2 } +\sigma^y )\otimes \mathbb{I}_{6\times 6}$, via the transformation $T = \exp\left(i\frac{\pi}{4}\sigma^{x}\right)\otimes \mathbb{I}_{6\times 6}$ , i.e. $\overline{G}_{km} = T^\dagger\cdot\widetilde{G}_{lm}\cdot T $, where:
 \begin{align}
     \left[\overline{G}_{km}\right]_{ab} = \frac{s_a\delta_{ab}}{i\omega_m-{s_a}\epsilon_{ka}},
 \end{align}
 where:
 \begin{align}
     s_a = \left\{
     \begin{array}{c}
     \,1, \text{ for }\, 1\leq a \leq 6 \,\text{, (particle sector)} \\
     -1, \text{ for }\, 7\leq a \leq 12\,\text{, (hole sector)} \,.
     \end{array}
     \right.
 \end{align}

 Hence, substituting Eq.(\ref{eq:gf1}) into  Eq.(\ref{eq:Pi_omega}), we obtain the following expression for the current-current correlation function:
 \begin{align}
    \Pi^{\mu\nu}_{\alpha\beta}&(i\omega_n) =  \nonumber \\ 
    & 2\sum_k \sum_{ab} [\overline{I}^\mu_\alpha(k)]_{ab}[\overline{I}^\nu_\beta(k)]_{ba}\, \chi^{ba}(k,\omega_n),
 \end{align}

 where $\overline{I}^\mu_\alpha(k) = T^\dagger\cdot S^{T}_k\cdot I^\mu_\alpha(k)\cdot S_k\cdot T$ is the flavor-resolved current matrix in the new basis and $\chi^{ba}(k,\omega_n) = -\frac{1}{\beta}\sum_m \overline{G}^{b}_{k,m} \, \overline{G}^{a}_{k,m-n}$. 
 
 The summation over the Matsubara frequencies can be carried out analytically and we have: 
 \begin{align}\label{eq:bubbles}
 \chi^{ba}(k,i\omega_n) & = s_as_b\frac{n_B(\overline{\epsilon}_{kb})-n_B(\overline{\epsilon}_{ka})}{\overline{\epsilon}_{ka} - \overline{\epsilon}_{kb} + i\omega_n}
 \end{align}
where $n_B(x) = 1/[\exp(\beta\, x)-1]$ is the Bose distribution function, and $\overline\epsilon_{ka} = {s_a}\epsilon_{ka}$.
Further simplifications arise if we consider the zero temperature limit of the expression in Eq.(\ref{eq:bubbles}), where $n_B(x) \sim - \theta(-x)$. Since $\epsilon_{ka} \geq0$, we finally have:
\begin{align}
    \Pi^{\mu\nu}_{\alpha\beta}(i\omega_n) =2\sum_{k}\sum_{ab}^\prime&\frac{\left[\overline{I}^{\alpha}_{\mu}(k)\right]_{ab}\left[\overline{I}^{\beta}_{\nu}(k)\right]_{ba} }{\epsilon_{ka} + \epsilon_{kb} + i\omega_n}\nonumber \,+\\  &
    \frac{\left[\overline{I}^{\alpha}_{\mu}(k)\right]^*_{ab}\left[\overline{I}^{\beta}_{\nu}(k)\right]_{ba}^* }{\epsilon_{ka} + \epsilon_{kb} - i\omega_n} ,
\end{align}
where $\sum_{ab}^\prime = \sum_{a\in \text{particle}}\sum_{b\in \text{hole}}$, 
and we used  $[\overline{I}^\mu_\alpha(k)]_{ab} =[\overline{I}^\mu_\alpha(k)]_{ba}^*$. 
Let us note, that the static limit of the current-current correlation function $\Pi^{\mu\nu}_{\alpha\beta}(i\omega_n = 0)$ is real, it only contribute to the singular part of the conductivity defined in Eq.(\ref{eq:curr_im_time}) 
\bibliography{ref_biblio.bib}